\documentclass[preprint,journal]{IEEEtran}
\usepackage{amsmath,amsfonts}
\usepackage{algorithmic}
\usepackage{algorithm}
\usepackage{array}
\usepackage[caption=false,font=normalsize,labelfont=sf,textfont=sf]{subfig}
\usepackage{textcomp}
\usepackage{stfloats}
\usepackage{url}
\usepackage{verbatim}
\usepackage{graphicx}
\usepackage{cite}
\usepackage{xcolor}
\usepackage{mathtools}
\usepackage{enumerate} 

\usepackage{float}
\DeclareMathOperator{\atantwo}{atan2}

\begin{document}

\title{Visualizing Routes with AI-Discovered Street-View Patterns}

 \author{Tsung Heng Wu, Md Amiruzzaman, Ye Zhao, Deepshikha Bhati, and Jing Yang
 \thanks{E-mail contact: twu10@kent.edu. \\ T-H. Wu, Y. Zhao, and D. Bhati are with the Department
 of Computer Science, Kent State University, OH, USA.\\ M. Amiruzzaman is with the Department of Computer Science, West Chester University, PA, USA.\\
 J. Yang is with the Department
 of Computer Science, University of North Carolina at Charlotte, NC, USA.}
 \thanks{\textcolor{blue}{This is the author’s version of the article that has been submitted to be published in IEEE Transactions on Computational Social Systems. The final version of this record is available at: 10.1109/TCSS.2024.xxxxxxx}}
 }

        % <-this % stops a space
% \thanks{This is the author’s version of the article that has been submitted to be published in IEEE Transactions on
% Visualization and Computer Graphics. The final version of this record is available at: 10.1109/TVCG.20xx.xxxxxxx}% <-this % stops a space
% % \thanks{Manuscript received April 19, 2021; revised August 16, 2021.}
% }

\markboth{IEEE Transactions on Computational Social Systems}%
{Authors version}


\maketitle

\begin{abstract}
Street-level visual appearances play an important role in studying social systems, such as understanding the built environment, driving routes, and associated social and economic factors. It has not been integrated into a typical geographical visualization interface (e.g., map services) for planning driving routes. In this paper, we study this new visualization task with several new contributions. First, we experiment with a set of AI techniques and propose a solution of using semantic latent vectors for quantifying visual appearance features. Second, we calculate image similarities among a large set of street-view images and then discover spatial imagery patterns. Third, we integrate these discovered patterns into driving route planners with new visualization techniques. Finally, we present VivaRoutes, an interactive visualization prototype, to show how visualizations leveraged with these discovered patterns can help users effectively and interactively explore multiple routes. Furthermore, we conducted a user study to assess the usefulness and utility of VivaRoutes.
\end{abstract}


\section{Introduction}
% \maketitle
Henry Miller, an American Author, said that ``One’s destination is never a place, but rather a new way of seeing things.'' In this paper, we present new computational and visualization methods that can help people ``see things'' along the rising-up roads. Roadside visual features reveal built environments and play a vital role in understanding a social system involving locations and geo-contexts. Thus, visualizing street views is of interest to applications such as urban and community planning, criminology, social equity, business and investment. Moreover, visualizing street views is of interest to personal route planning, since they link to many social, economic, and environmental factors that can affect personal route decisions. For example, some people may prefer to drive in greenery while others may want to navigate in urban surroundings.

We are all familiar with the visualizing routes on maps, mostly as color-coded trajectories, to explore, select, and navigate to destinations. Apparently, visualization of street views can be an elegant complement to existing tools. Unfortunately, street-view information is often represented by a large set of spatially sampled and heterogeneous pictures. Directly visualizing them together with urban structures and maps can easily lead to visual clutter and thus overwhelm users. 

To make street-view visualization compatible with route views and enable easy understanding, it is mandatory to find a summarized way to present street-view images. According to ``pattern theory'', a theoretical model in visual analytics \cite{andrienko2022seeking}, abstraction in data analysis is achieved by finding patterns in data distributions. The model also regards pattern discovery as a fundamental operation in visual analytics processes. In this paper, we propose several computational approaches to discovering street-view patterns. Machine Learning (ML) tools are employed for data transformation to handle the big size and diversity of the raw images. Afterward, we develop new visualization methods to integrate these visual patterns within an interactive interface. The problems we tackle and our technical approaches include:

\begin{itemize}
\item \textbf{Finding quantitative representation to compute similarities among street-view images:}  
It is important to define what are similar styles of street views. There exists no simple equation and optimal solution, but rather an issue related to human perception and experiences. We explore several Deep Learning (DL) methods to compute quantitative vectors in latent spaces that can represent inherent imagery features.  

\item \textbf{Extracting area-aware visual patterns from street-view images:} With the latent vectors, we further employ several clustering methods to discover a small group of ``visual appearance patterns'' (VaPatterns). These clusters provide a succinct set of the aimed ``patterns''. 

\item \textbf{Visually exploring the visual patterns over routes:} 
A set of deliberately designed visualizations present the discovered VaPatterns together with roads and geographical context. Visual interactions enable users to quickly explore different routes and compare them. The system also supports multi-resolution exploration with both coarse and fine details over different parts of routes. A visualization prototype, named as \textbf{VivaRoutes} (\textbf{Vi}sualizing \textbf{v}isual \textbf{a}ppearance of Routes), is implemented. The system can be combined with existing route planners (currently Google Map is used).
\end{itemize}

The main contributions of this paper can be summarized as:
\begin{itemize}
    \item We identify the importance of street-level visual appearance and include street-view features in routing services. 
    \item We discover and quantify visual appearance features by employing DL and ML tools, which enable efficient visualization and interactive exploration. 
\item A visualization system, VivaRoutes, is developed for visualizing and comparing visual appearances on routes.
\item A few case studies have been conducted to illustrate the usefulness and effectiveness of VivaRoutes. 
\end{itemize}

VivaRoutes can be used for planning and exploring routes by tourists and commuters. Urban planners and community workers can also use it for their community study. Our approach has the potential to improve visualization systems in fields such as tourism, urban planning, and the social-economical study of communities.

\begin{figure*}
    \centering
 \includegraphics[width=\textwidth]{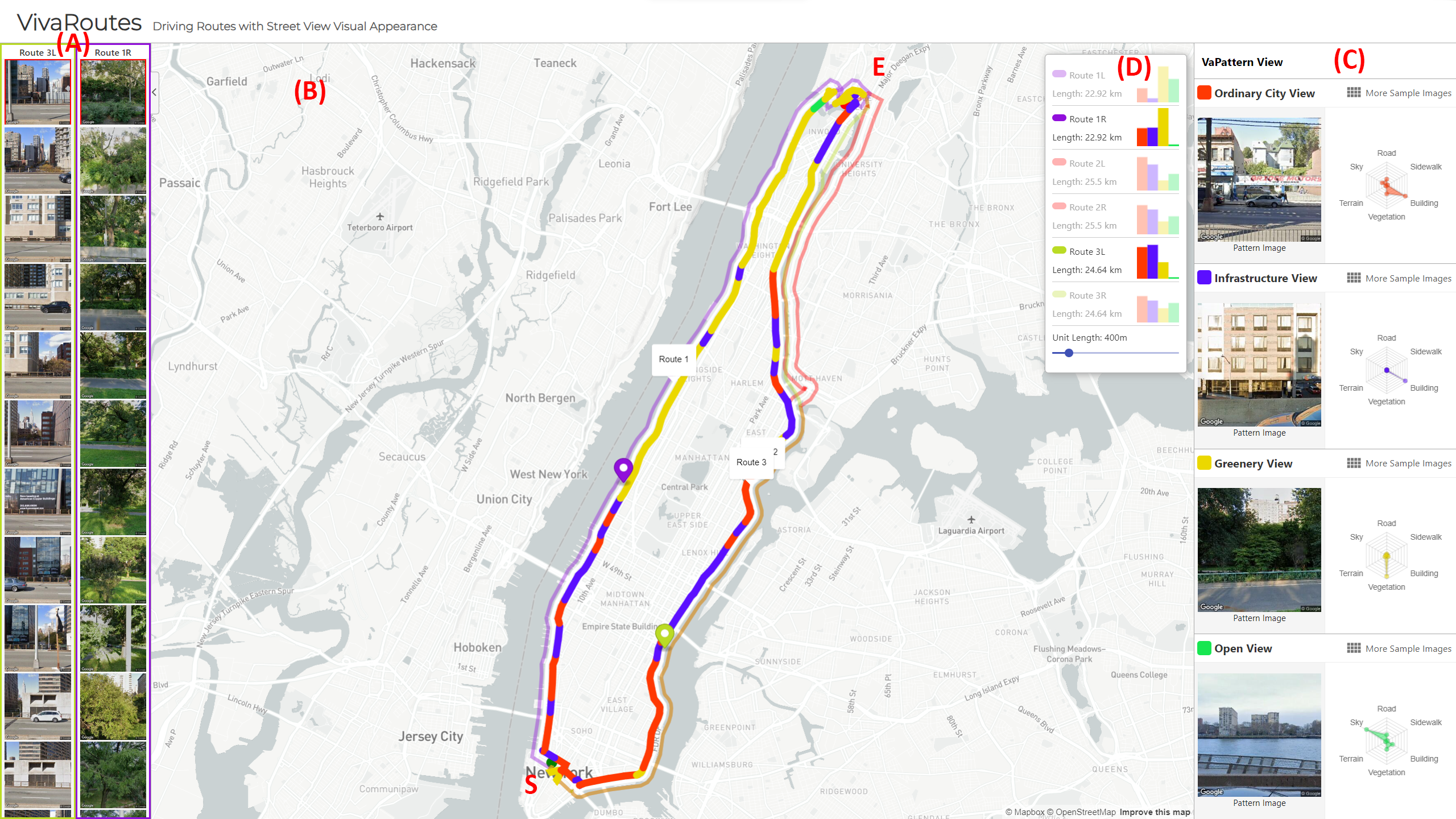}
\vspace{-15pt}
\caption{VivaRoutes interface for visual exploration of driving routes in Manhattan, NY. (A) Route image window shows the captured street-view images (extracted from Google Street View API) on two selected routes from a starting location (S) to a destination (E). (B) Map view with the two routes colored according to discovered VaPatterns. (C) The VaPatterns are visualized to show the street-view patterns. (D) A map inlet for route information, VaPattern distribution, and user control. }  \vspace{0pt} \label{fig:main_interface}
\end{figure*}
% \begin{figure*}
%     \centering
%  \includegraphics[width=\textwidth]{images/Overview_w_segment.png}
% \vspace{-5pt}
% \caption{The main interface of the VivaRoutes system. (A) segment visualization panel, (B) individual map for selected routes, (C) Cluster visualization panel to provide overview of the scenic groups, (D) global map view, (E) Route visualization panel. Please see details in Sec. \ref{sec:vivaroutes} and Sec. \ref{sec:case_studies}.
% } \vspace{-5pt} \label{fig:main_interface}
% \end{figure*}

\section{Related Work}

\subsection{Geographic Routing}
Route planning is a research topic in transportation and urban studies, as well as in logistics, autonomous vehicles, and energy saving. 

A routing algorithm calculates paths between two locations (e.g., source and destination) \cite{nha2012comparative} with different metrics \cite{fawcett2000adaptive} such as distance, cost, tolls, and time \cite{shoaib2013traffic}. Context-aware route planning further adds crime, energy, and social information to route computation \cite{malakooti2006multiple,musolesi2005adaptive} to find an optimal and feasible set of routes. Several studies (e.g., \cite{samet2008scalable,sankaranarayanan2010query,yang2015toward,guo2020context}) combine multiple criteria into one decision metric, such as weighting crime rate together with distance criteria. Moreover, recent studies of personalized route recommendations use crowdsourcing and/or social media data in finding optimal routes for specific users \cite{chen2014tripplanner,mirri2014combining,prandi2014mpass}. User experiences and preferences are mined and used in route computation. For example, Mirri et al. \cite{mirri2014combining} find routes that provide more accessibility for elderly people by collecting data from Foursquare and Yelp. The street-view imagery, which is the focus of this paper, can add another dimension to route study and give users new options in route decisions.  

\subsection{Street-view Imagery}
Street-view visual contents, such as different views of roads, buildings, greenery, sky openness, etc., form an important environmental and social factor, which has become a critical research topic in landscaping, urban planning, transportation, and social studies \cite{shen2017streetvizor,yatmo2008street}. Traditional approaches are often conducted by in-person surveys, mapping, and remote sensing of the built environment \cite{shen2017streetvizor}. Many researchers have used Google Street-View (GSV) images in community studies. For example, assessing damage made by tornadoes \cite{zhai2020damage}, understanding the association between the built environment and health outcome \cite{nguyen2019using}, finding green areas \cite{berland2017google}, and discovering criminal activities \cite{gerell2017violent,vandeviver2014applying} and animal habitats \cite{olea2013assessing}. 

The recent developments in computer vision and DL technologies have made this process less expensive and faster. They can find objects and extract semantic categories from street-view images and videos (e.g., Segnet \cite{badrinarayanan2017segnet} and PSPnet \cite{zhao2017pyramid}). Several studies have taken advantage of the DL models to extract semantic categories from images and use them in social studies (e.g., \cite{zhou2019social}, \cite{ning2021sidewalk}. In this paper, we extract visual appearance features with DL tools for route information visualization.

\subsection{Spatial and Street-view Data Visualization}
Various visualization systems have been developed to make sense of geospatial data in transportation and urban applications \cite{andrienko2013visual,andrienko2017visual,von2012visual}. In particular, street-view images are utilized in a few VA systems \cite{shen2017streetvizor,li2017building,Zeng2018} for visual comparison of spatial distributions and exploration of fine-grained visual details at the street scale. Geo-narratives integrate opinions and descriptions with geo-videos for social geographical research \cite{jamonnak2020geovisuals}. In this paper, we extract and design visualizations of visual appearance features with the new goal of providing information on multiple driving routes.

\section{Overview: Rationale and Methodology} \label{sec_overview}
Our goal is to add visual information about the street-side landscapes to route visualizations. The landscape's visual appearances can be captured from a large set of street-view images. However, these images need to be summarized as visual appearance patterns for easy visualization and understanding. It defines our computational goal: 

The patterns would preferably match our mental pictures that represent our mind's experiences of perceiving street scenes. Usually, the mental pictures are derived with high-level abstraction and categorization. For instance, we consider one section of a road with open and green views as ``country style'', and another section with mixed building and road views as ``town style''.

We then seek ML techniques to discover the patterns from street-view images. This includes two major computational tasks (\textbf{C1-C2}):
\begin{itemize}
\item \textbf{C1: Defining similarities among street view images}: We need to find quantitative representations of the images so that ``distance'' among different images can be computed. This similarity should be able to reflect the typical visual perception difference from human observers;
\item \textbf{C2: Discovering patterns by clustering the images}: We need to use appropriate clustering methods and form a necessary number of clusters. These clusters discover the patterns that can represent the perceiving styles of observers.
\end{itemize}
In Sec. \ref{sec:patterndiscovery}, we show our exploration of multiple computing algorithms in defining similarities among the images and clustering these images. 

To explore the visual patterns discovered, we further develop VivaRoutes. It is an interactive route visualization interface that integrates new visualizations and the computing algorithms in (Sec. \ref{sec:vivaroutes}). The design goal of its visualizations is to allow people to explore alternative street-level visual appearance features along candidate driving routes. The interface should be easy to use and interactive. We identify several visualization design tasks (\textbf{V1-V4}) including:
\begin{itemize}
    \item \textbf{V1: Visualizing street-view patterns intuitively}: We need to summarize the extracted street-view patterns and allow users to easily understand them. The patterns need to be added to existing map views of driving routes. The new information should fit well into the geographical context and not add more recognition burden for users. 
    
    \item \textbf{V2: Supporting interactive route exploration based on street-view information}: We need to design new visualizations and interactions to help users interactively explore and understand route features and examine their road views. 

    \item \textbf{V3: Facilitating easy comparison of alternative routes with street-view information}: We need to provide a solution for route comparison through street-view patterns, so that users have a new way to examine and select routes.
    
    \item \textbf{V4: Visualizing street-views in multiple scales}: We need to allow users to study the visual appearance patterns in fine scales (i.e., specific areas or street segments) on routes. The study should also be combined with views of street-view images.
\end{itemize}

\begin{figure}
    \centering
 \includegraphics[width=\columnwidth]{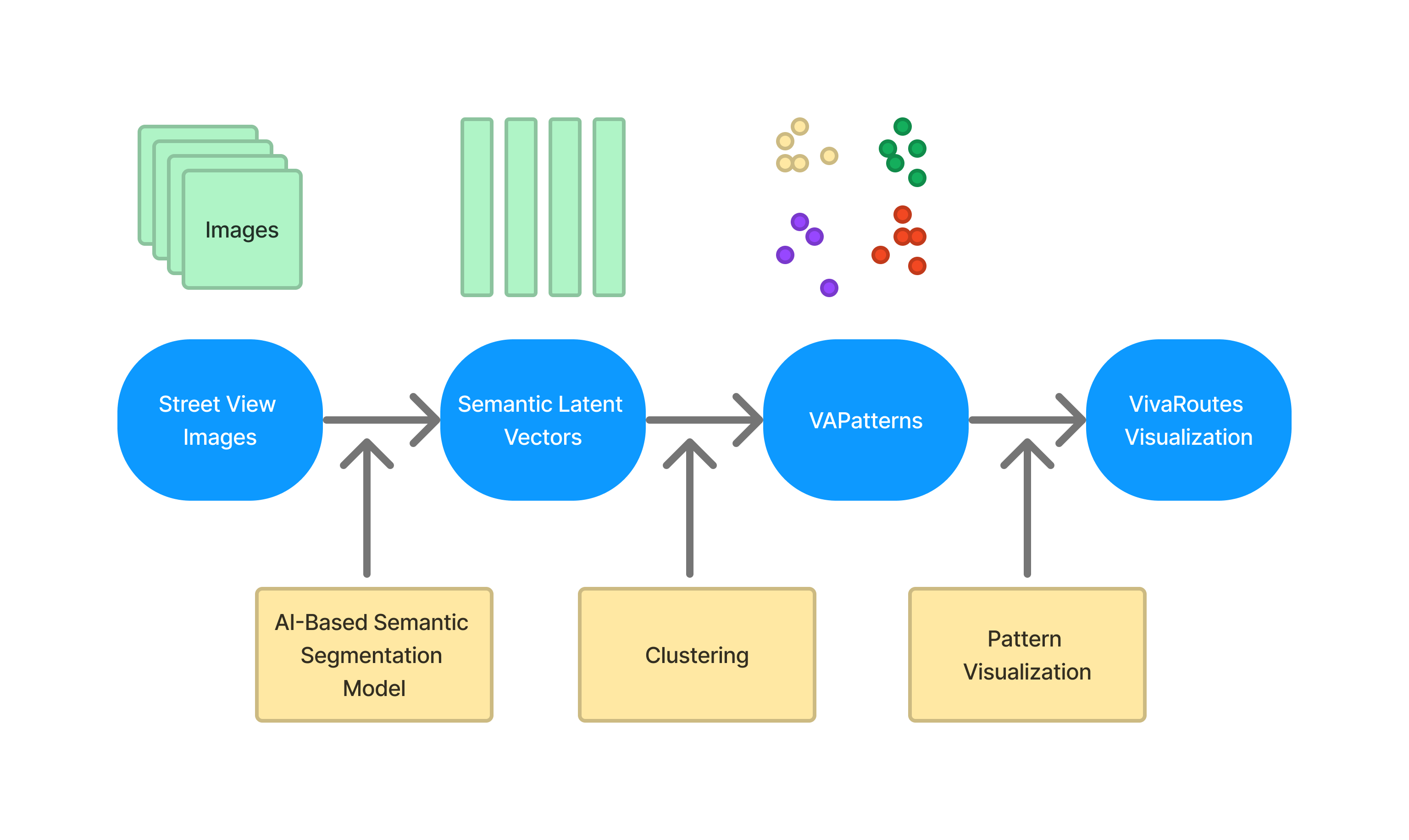}
\vspace{-25pt}
\caption{Illustration of the workflow of VaPattern extraction and visualization. }  \vspace{-10pt} \label{fig:flow_chart}
\end{figure}

Fig. \ref{fig:flow_chart} illustrates the workflow of VaPattern extraction and visualization. Through an AI-based semantic segmentation model, street view images are represented by semantic latent vectors. These vectors enable the use of clustering methods to group the images into multiple clusters, i.e., VAPatterns. These patterns are then visualized in an interactive map-based visualization system.

Next, we introduce our data collection process, explore multiple AI methods for pattern extraction, and present our new visualizations and interface that fulfill the specified requirements.

\section{Data Collection}
\label{sec:datacollection}
We acquire street geometry data and street-view images from public sources. There exist many route planner APIs that compute and suggest several alternative routes from a given source to a destination location, such as Open Source Routing Machine (OSRM), Mapbox, and Google Directions. In this paper, we used Google Directions API which recommends three alternative routes.

Next, these routes are matched to street segments, which are then used in both street-view image retrieval and visualization on maps. The road network geometry data of a selected city is downloaded from the open GIS data repository, OpenStreetMap (OSM) \cite{bennett2010openstreetmap}. 

Along these routes, we acquire street-view images from Google Street-View (GSV) with the available public API \cite{amiruzzaman2021classify,amiruzzaman2023ai}. In the implementation, we need to calculate the heading direction of each street segment on a route. Then this direction is used to compute the view direction angles toward the left side and right side of the street. This is an essential step as the important visual appearance features are on street sides, but not on the forward (backward) views on the road itself \cite{amiruzzaman2021classify}.

To refine the accuracy of this computation, for each street segment (which may be a curve), we divide it into small chunks each having a length of about 20 meters. Then the heading direction of the chunk is computed as: 
\begin{eqnarray}\label{eq:heading}
\theta &=& \atantwo(x,y) \\
\text{where} \nonumber \\
x &=& \cos({\phi}_1)\times\sin(|{\lambda}_1-{\lambda}_2|), \nonumber \\
y &=& \cos({\phi}_1)\times\sin({\phi}_2) \nonumber \\ &&-\sin({\phi}_1)\cos({\phi}_2)\times\cos(|{\lambda}_1-{\lambda}_2|). \nonumber 
\end{eqnarray}
Here, ($\phi_1$, $\lambda_1$) and ($\phi_2$, $\lambda_2$) are the latitudes and longitudes of the start and end point of this chunk, respectively. After finding the heading direction (i.e., $\theta$) of each chunk, we get the view angles towards the left side and right side. Then, left-side and right-side street-view images at the mid-location of each chunk are retrieved from GSV API by providing the latitude, longitude, and view angles. Therefore, we acquire the visual appearance images for all roads in a spatial region, with a spatial resolution of 20 meters. This resolution may be adjusted for a balance of accuracy and computational load.

% \begin{verbatim*}
% https://maps.googleapis.com/maps/api/streetview?size=400x400&location={0},{1}&fov=80&heading={2}&pitch=0
% &key=YOUR_API_KEY.
% \end{verbatim*}

Raw street-view images need to be processed to identify visual objects inside them. This is a process of image-based semantic segmentation that extracts street-view object categories for geographical scenes. Moreover, these categories on multiple images in a region are used to retrieve higher-level semantic features for meaningful and intuitive representations. We employ state-of-the-art AI and ML algorithms to address these tasks.  

\begin{figure}[h]\centering
 \includegraphics[width=1\columnwidth]{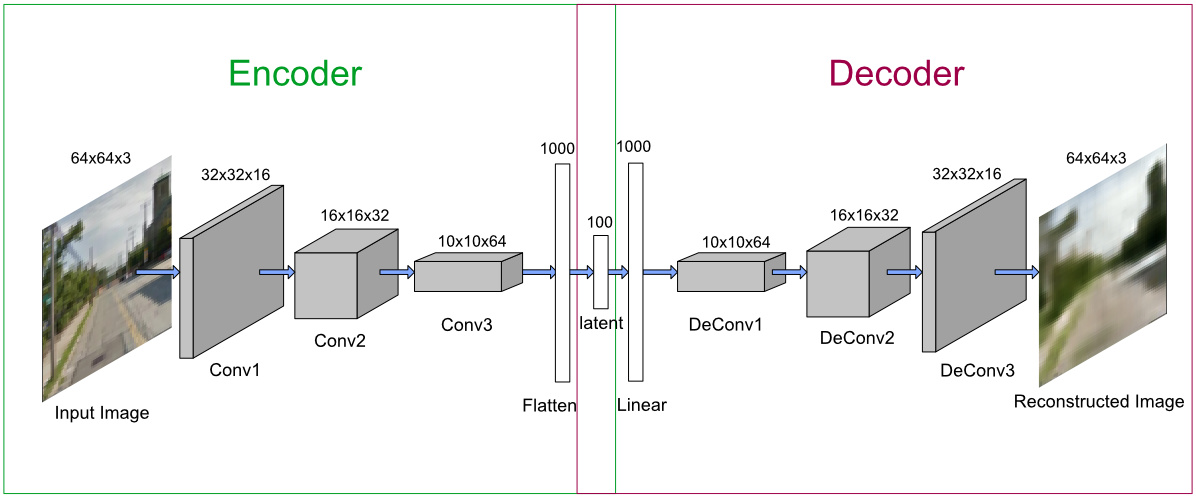}\vspace{-5pt}
    \caption{ The architecture of our autoencoder model. }\vspace{-5pt}
    \label{fig:encoder}
\end{figure}

 \section{Visual Appearance Pattern and Computing Methods} \label{sec:patterndiscovery}

\subsection{Visual Appearance Patterns}
In real world scenarios, individuals often perceive visual appearances within a built environment as abstract and comprehensive patterns. These ``patterns'' are distilled from a large set of views by human knowledge and experience. They inherently link human perception impressions with urban settings. For example, one pattern may relate to most greenery and open sky, leading to ``country style views''. Another pattern may be associated with most buildings and sidewalks, leading to ``downtown style views''. These patterns may be disparate for different locations due to terrain, weather, architecture, etc. 

In a given geo-region, we can discover such patterns from the street-view images by a visual appearance clustering approach. Ideally, if we can identify a small group of patterns that capture the visual appearances of the street views well, we will be able to employ them for effective visualization of driving route features. We call these patterns VaPatterns.

The major challenge in addressing this task is the diversity of the street-view images, influenced by factors such as the setups and qualities of the cameras, driving conditions, weather, etc. Pixel-based image clustering methods cannot be directly applied. We thus seek help from DL techniques, where the task is divided into two steps: (1) defining quantitative vectors for image similarity computation, and (2) clustering the street-view images into multiple patterns. Next, we show our explorations of multiple approaches.  

\subsection{Calculating Street-View Image Similarity}

We need to find a quantitative representation of each input streetview image. The representation should reflect the stylish similarity and disparity among the given images so as to discover the patterns. We explore three different encoding methods from recent DL techniques to find quantitative vectors to encode the images. Next, we describe and discuss our experiments with these methods to find an optimal solution to our application.

\begin{figure}[ht]
\centering
\includegraphics[width=1\columnwidth]{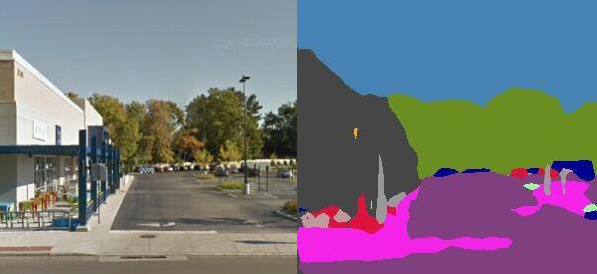}\vspace{-5pt}
\caption{A street-view image (left) extracted from Google Street View API and its visual semantic categories extracted by PSPNet (right). The pixels are colored by their categories, such as blue for the sky, grey for buildings, etc. The category distribution is presented in Table \ref{tbl:semantic_vector}. }\vspace{-5pt}\label{fig:gsv_image}
\end{figure}

\subsubsection{Using Autoencoder for Image Encoding}
We design an autoencoder to encode street-view images in a 100-dimensional latent space. The model structure is illustrated in Fig. \ref{fig:encoder}. In training, a group of sampled street-view images in a geographical region passed through three convolutional layers and two linear layers to extract the relevant information (i.e., the information makes them distinguishable), and then those were flattened to get the latent vectors. The vectors are further used to reconstruct the street-view image. The activation functions between layers are mostly the Rectified Linear Unit (ReLU) function, except for the output of the encoder, where we use the hyperbolic tangent activation function to normalize the latent vector to [-1, 1]. The activation function of the output of the decoder is sigmoid because it has to map the pixel values back to [0, 1]. In training, the loss function is implemented as the pixel-wise difference between the reconstructed image and the input image. From GSV, the original street image size is $300\times300$ with three color channels. They are down sampled and resized to $64\times64$ for the autoencoder to achieve fast and effective training.

This architecture is compact yet efficient for the $64\times$64$\times3$ input size. Because our purpose of using the autoencoder is to find a low-dimensional representation rather than image content generation, we do not choose the more complex variations of the autoencoder.

\begin{figure*}[t]\centering
 \includegraphics[width=0.3\textwidth]{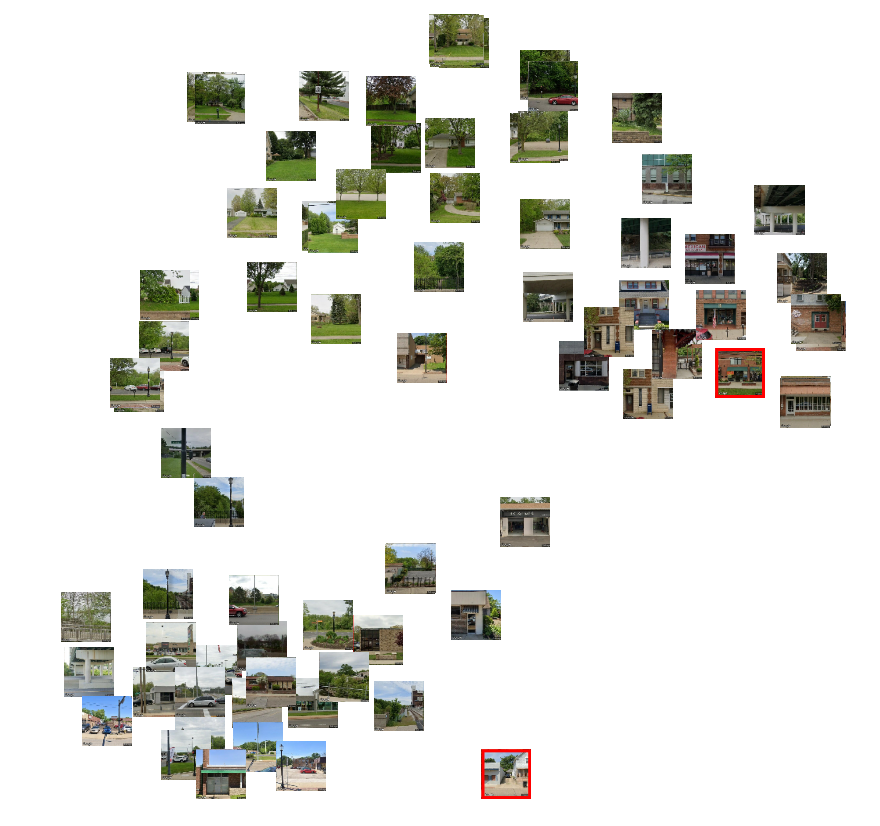}
 \includegraphics[width=0.3\textwidth]{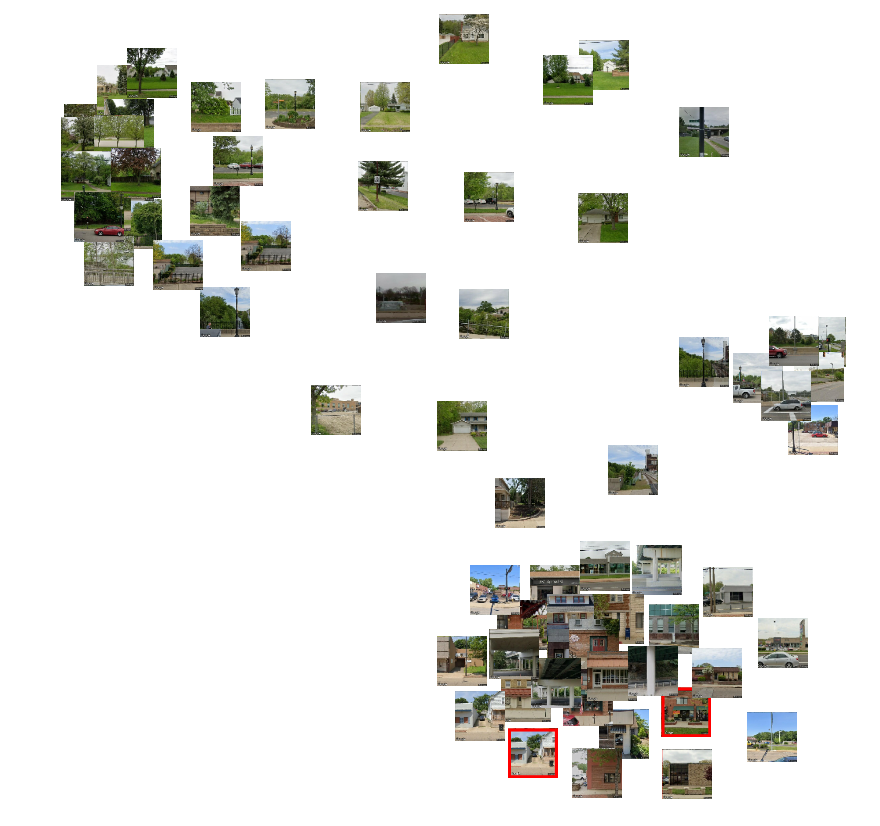}
 \includegraphics[width=0.3\textwidth]{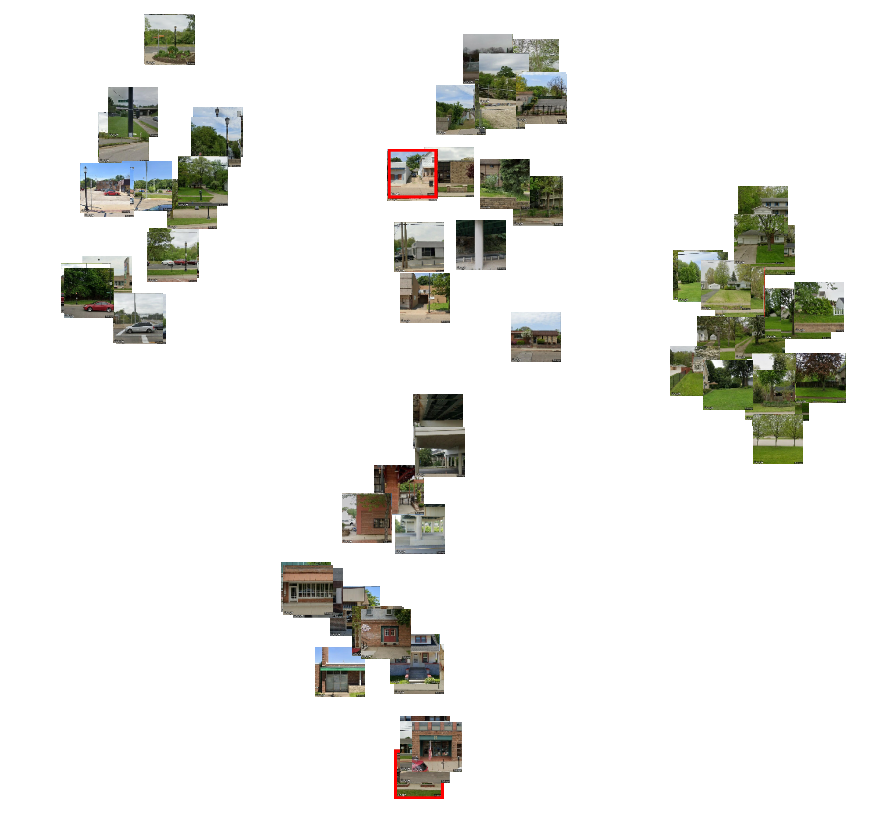}\vspace{-5pt}
    \caption{ Using different image encoding approaches for image similarity for a diverse set of sample streetview images extracted from  Google Street View API. (Left) Using autoencoder; (Middle) Using semantic categories; (Right) Using semantic latent vector;} \vspace{-5pt}
    \label{fig:imageencodingexample}
\end{figure*}

\begin{figure}[t]\centering
 \includegraphics[width=0.48\columnwidth]{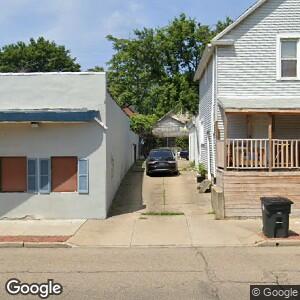}
 \includegraphics[width=0.48\columnwidth]
 {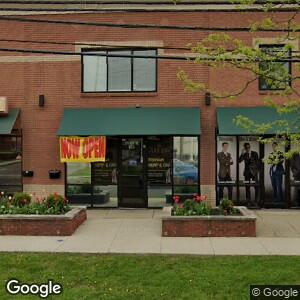}
 \vspace{-5pt}
    \caption{ Two images extracted from  Google Street View API in a Midwestern geo-region that represent two different visual patterns. (Left) residential area view; (Right) business district view; }\vspace{-5pt}
    \label{fig:exampleimage}
\end{figure}

\subsubsection{Using Semantic Categories for Image Encoding}
Recently, DL models largely improve the effectiveness and accuracy in extracting visual objects and meaningful categories from natural images, which have been used to analyze and explore fine-grained information from street-view images \cite{amiruzzaman2021classify,shen2017streetvizor,Zeng2018}. We employ PSPNet \cite{zhao2017pyramid}, a popular deep neural network model, for extracting visual semantic categories. The PSPNet outputs a 19-dimension vector for an input image while each vector component represents the percentage of pixels belonging to one category in 19 different visual categories. They are road, sidewalk, building, wall, fence, pole, traffic light, traffic sign, vegetation, terrain, sky, person, rider, car, truck, bus, train, motorcycle, bicycle \cite{zhao2017pyramid}. An example is illustrated in Fig. \ref{fig:gsv_image}, and the corresponding result vector is shown in Table \ref{tbl:semantic_vector}.

\begin{table}[h]
\centering
\caption{A sample vector obtained from semantic segmentation from a street-view image in Fig. \ref{fig:gsv_image}} \label{tbl:semantic_vector}
\begin{tabular}{|l|l|l|l|l|}
\hline
Road   & Sidewalk  & Bicycle & $\cdots$   & Building   \\\hline
0.31 & 0.03   &  0 & $\cdots$          & 0.18   \\\hline      
\end{tabular}
\end{table}

However, some categories may not be significant for our purpose (e.g. bicycle). We can thus reduce the dimensionality of the vectors. Toward this goal, we collect a large number (about 50,000) of street-view images in each of the US regions being studied in VivaRoutes. By studying the percentages for each semantic category in all the images in one region, we identify six major categories - road, sidewalk, building, vegetation, terrain, and sky. To justify this primary observation, we conduct a \textit{dispersion measure analysis}. It is often used in ML to reduce the dimensionality of high-dimensional data and select important features \cite{ferreira2012efficient}. It is shown to be effective in extracting features from both nominal and categorical data \cite{roy2019dispersion}. It computes a dispersion ratio where a high value indicates a high-relevance feature and a low value links to a low-relevance feature in the given dataset. For our study, we use a cutoff threshold of 1.0 to select the top six categories. The results are the same as our direct selection. These categories form a six-dimensional vector for each image encoding as:
\begin{equation} \label{eqn:sixcategories}
    V = \{Road, Sidewalk, Building, Vegetation, Terrain, Sky\}.
\end{equation}

%Please note that the effective representative categories may be reliant on spatial locations. This dimension analysis can be applied to extract optimal categories specifically for different areas.

\subsubsection{Using Semantic Latent Vector for Image Encoding}
In addition to direct use of the output semantic categories from the semantic segmentation neural network, we take a deeper look into their architecture.

We realized that the latent vector from the encoder cipher the semantic information and thus can be utilized in the similarity computation. It uses a higher-dimensional vector and thus potentially includes more latent semantic features than the final output with 19-dimensional categories.  

In computation, we use a 1000-dimensional latent vector (semantic latent vector) in the ResNet-50 backbone of the DeepLab V3 \cite{deeplabv3} semantic segmentation model. The model was trained using the Cityscapes dataset, which is best suited for our use cases.

\subsubsection{Results and Discussion}
To show the effects, we select a diverse set of images as an example. It contains different street views in a Mideast region. After mapping these images to three different encoding vectors with the three approaches, these images are visualized in a t-SNE view in Fig. \ref{fig:imageencodingexample}. In this figure, the left view is from the autocoder, the middle view is from the semantic category vector, and the right view is from the latent vector in the semantic segmentation network. First, it can be seen that the semantic latent vector method can group the grassland images closely together. But the other two methods cannot group them very closely. For instance, two grassland pictures give users the same visual feelings but in semantic categories, the percentages of the sky in them are quite different so their representative vectors are not considered similar. Second, two pictures in Fig. \ref{fig:exampleimage} show two images that have different visual styles. The left one is more like a dense residential view, while the right one presents a small-town business view. These two images are highlighted in Fig. \ref{fig:imageencodingexample}. 

It can be seen that the autoencoder separates them, but one of them is identified as an outlier at the bottom. The semantic category vector method groups them and other similar pictures together. In contrast, the semantic latent vector method separates them and puts them in the vicinity of similar views. From these observations, we found that using the 1000-dimensional semantic vectors to encode images can better represent them for our purpose. A possible explanation is that the autoencoder can detect low-level features but does not consider semantic features, while the semantic categories discover semantic categories but do not include enough low-level features. The semantic latent vectors however can capture both low-level and semantic features since it is from a pre-trained semantic segmentation neural network. Therefore, we use the semantic latent vector for the following pattern discovery work.

\subsection{Discovering Visual Appearance Patterns}
\label{sec:clustering}

ML clustering algorithms are employed to group street-view images into VaPatterns. The appropriate number of clusters (i.e., the number of ``patterns'') may not be a fixed value. For example, in some areas, there may only be 1 or 2 different visual appearance styles while other areas may have more styles. This depends on the size and social factors of the region. 

We study several clustering methods to find a good solution in our experiment datasets. These methods include the supervised k-means and agglomerative hierarchical clustering, where varying numbers of clusters are tested, and an unsupervised meanshift clustering method. They are applied to all sampled images from specific geographical regions (e.g. 5000 images in the Midwestern area).

In our experiment, we found that the meanshift methods cannot successfully group the images into meaningful clusters. Thus we adopted the supervised methods and tested different numbers of clusters. The selection of an optimal approach is guided by the Silhouette score, which is a popular metric for clustering result evaluation \cite{dinh2019estimating,reynolds2006clustering}.

In our work with the GSV street-view images in the Manhattan area, New York, the Silhouette score drastically drops when $k$ = 4 changes to $k$ = 5. Therefore, we choose the four clusters that resulted from the hierarchical clustering as VaPatterns. They represent four typical visual appearance patterns in this geographical area. Please note that this work only needs to be done once for a given geographical area in the pre-processing stage.

\begin{figure*}[t]
\centering
\includegraphics[width=1.0\textwidth]{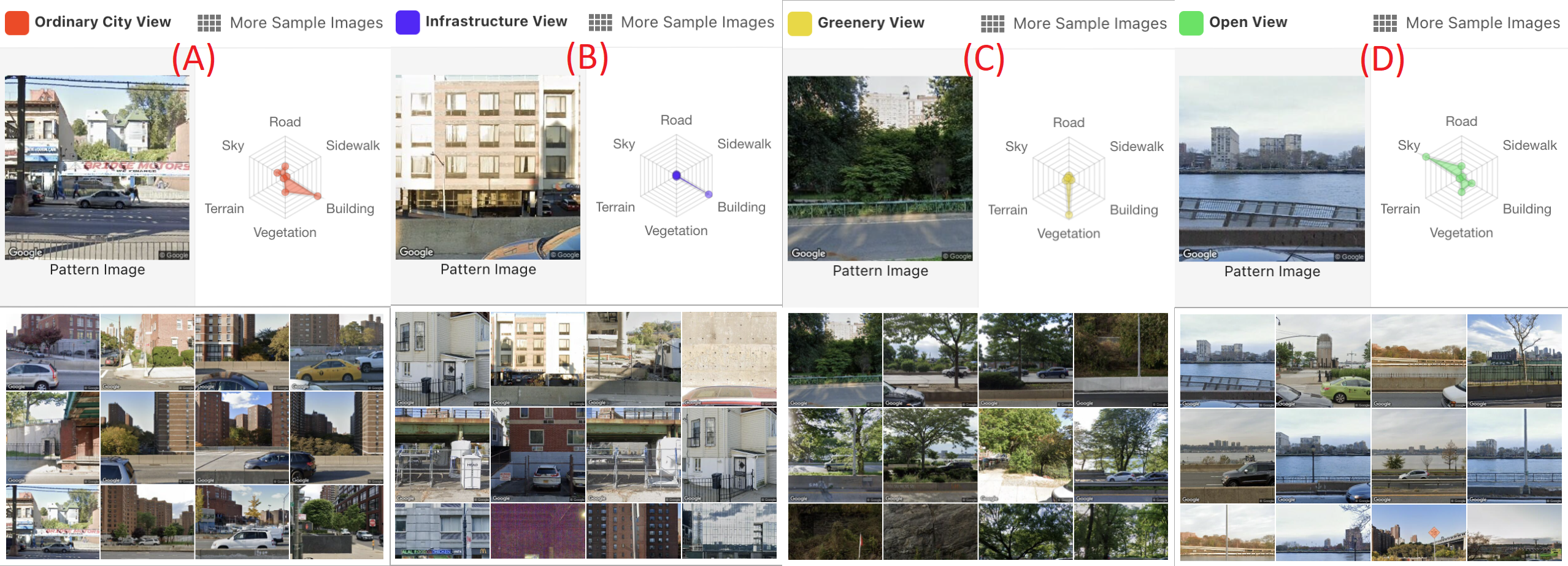}
\caption{Discovering Manhattan’s visual appearances patterns with four VaPatterns from (A) to (D). The pattern image and a group of sample images (extracted from Google Street-View API) are shown, together with the radar chart that visualizes the semantic category distribution.}  \label{fig:VaCluster}
\end{figure*}

\section{Visualizing Streetview Patterns}
\label{sec:vivaroutes}

To meet the visualization requirements (V1-V4) in Sec. \ref{sec_overview}, we design several new visualizations integrating the extracted street-view patterns. The new visualization interface of VivaRoutes integrates these new visualizations with existing route map view, as illustrated in Fig. \ref{fig:main_interface}.

\subsection{Design Overview}
\noindent \textbf{New Perspective of Route:} Our goal is to visualize street-level views along roads. Street views are different on the left and right sides. Therefore, in our system, each traditional route from origin to destination should be considered as two distinct trajectories, denoted as route L (left) and route R (right), and displayed as a pair of trajectories. To increase flexibility and easy comparison, we allow users to select, highlight, and compare two different trajectories, such as Route 1R with Route 3L. For simplicity, in the following, we sometimes refer to a route indeed representing one route trajectory of a specific side.

Our visualization designs are summarized below according to the requirements including: 
\begin{itemize}
\item \textit{For V1:} The discovered VaPatterns are abstracted from various street views. We design a \textbf{VaPattern view} to display their semantic contents so that users can quickly understand their key visual components such as open sky, greenery, etc.

\item \textit{For V2 and V3:} We include street pattern visualizations into a traditional map view of driving routes in the \textbf{VivaRoutes interface}. It provides an additional dimension for users to study candidate routes from their origin to destination locations. Users can also study and compare routes with VaPatterns through a map inlet.

\item \textit{For V4:} Users are allowed to explore routes with VaPatterns with zoom-in and out operation. Users can also drag markers on the map, and compare route details with their street-view images.
\end{itemize}

These views are integrated into the VivaRoutes interface as illustrated in Fig. \ref{fig:main_interface}: 
\begin{itemize}
    \item VaPattern View (Fig. \ref{fig:main_interface}{C}) which illustrates VaPatterns to quickly present their representatives of visual appearance patterns. 
    \item Map View which visualizes the candidate driving routes with VaPatterns in their geographical context (Fig. \ref{fig:main_interface}{B}). 
    \item A map inlet (Fig. \ref{fig:main_interface}{D}) where a bar chart shows the VaPattern distribution over each route. It can support users to quickly find the visual appearance information on each side of routes, and also compare them quantitatively. Users can select and highlight two routes for a visual comparison. 
    \item The street-view images (Fig. \ref{fig:main_interface}{A}) of the two highlighted routes. They are shown based on user interaction by dragging markers (arrows) on the map. It is important that users can examine raw street-view images together with the route view.

\end{itemize}
Next, we describe the visualization design and functions in detail.

\subsection{VaPattern Visualization}
\label{sec:VaCluster}

We visually display VaPatterns to reveal visual features along routes transparently and intuitively. To display the inherent contents of VaPatterns in a way that is easily understandable and accessible for users, we represent VaPatterns using their semantic information. In particular, each image belonging to one pattern has a six-dimensional vector as described in Eqn. \ref{eqn:sixcategories}. We compute a representative \textbf{pattern vector} by averaging such vectors of all the images. Moreover, a \textbf{pattern image } is retrieved by finding a street-view image whose semantic categorical vector is the closest to the pattern vector.  

Afterward, in the VaPattern view (Fig. \ref{fig:main_interface}{C}), a radar chart (\cite{albo2015off}) is adopted to present the pattern vector with its distribution in the six semantic categories (see Fig. \ref{fig:main_interface}{C}). Observers can easily find major visual categories (Eqn. \ref{eqn:sixcategories}) within the corresponding pattern. The corresponding pattern image is shown for direct understanding. Users can open a matrix view to explore more raw street-view images with this pattern.  Moreover, users can name the patterns with preferred names such as ``Ordinary City View'', ``Infrastructure View'', etc. Each pattern is assigned a unique color which is also used in the map and other views for coordination. 

\subsection{Integrating VaPatterns with Route Visualization}
People are familiar with a map-based interface where multiple candidate (recommended) routes are visualized over map view. Therefore, we propose to display the VaPattern features over these routes within these interfaces, in a way that respects the established cognitive framework of users and does not add extra mental burdens to them. Our basic approach is mapping the VaPatterns to different colors and showing them on the trajectories. In addition, we draw two trajectories for each route representing the left and right sides, respectively. 

There exists a key technical challenge: we cannot visualize the VaPatterns in different colors at all image sampling locations along a road. This will cause overwhelming color variation and cannot provide multiresolution control. Therefore, the visualization is implemented based on route segments with the following algorithm: First, the route is divided into ``segments'' with a user-adjustable length (on Fig. \ref{fig:main_interface}{D}). The segments make it possible for controllable multi-resolution information convey. Then, the street-view images of each segment are retrieved. These images belong to different VaPatterns, but the dominant one, representing the majority of images, is used to show the pattern of this route segment. This dominant VaPattern provides the color of the segment on the map. We use color-coded line segments on the map (Fig. 1E) to show the patterns, allowing viewers to easily identify patterns directly on the map. Users can choose preferred colors for different patterns.

To observe the VaPattern distribution, users can interactively select two route trajectories and highlight them on the map inlet (Fig. \ref{fig:main_interface}D). The bar chart gives users a summary of VaPattern distribution and the route length. Here, Route 1R and Route 3L are selected and highlighted with their changing patterns. Users can also drag two markers on them for drill-down study.

In the route image window (Fig. \ref{fig:main_interface}A), users can choose which side of the routes to be shown in the panel. The raw images around the marker locations are presented. In this window, users can select to show more than 2 columns for their study.

\begin{figure*}[t]
\centering
\includegraphics[width=1.0\textwidth]{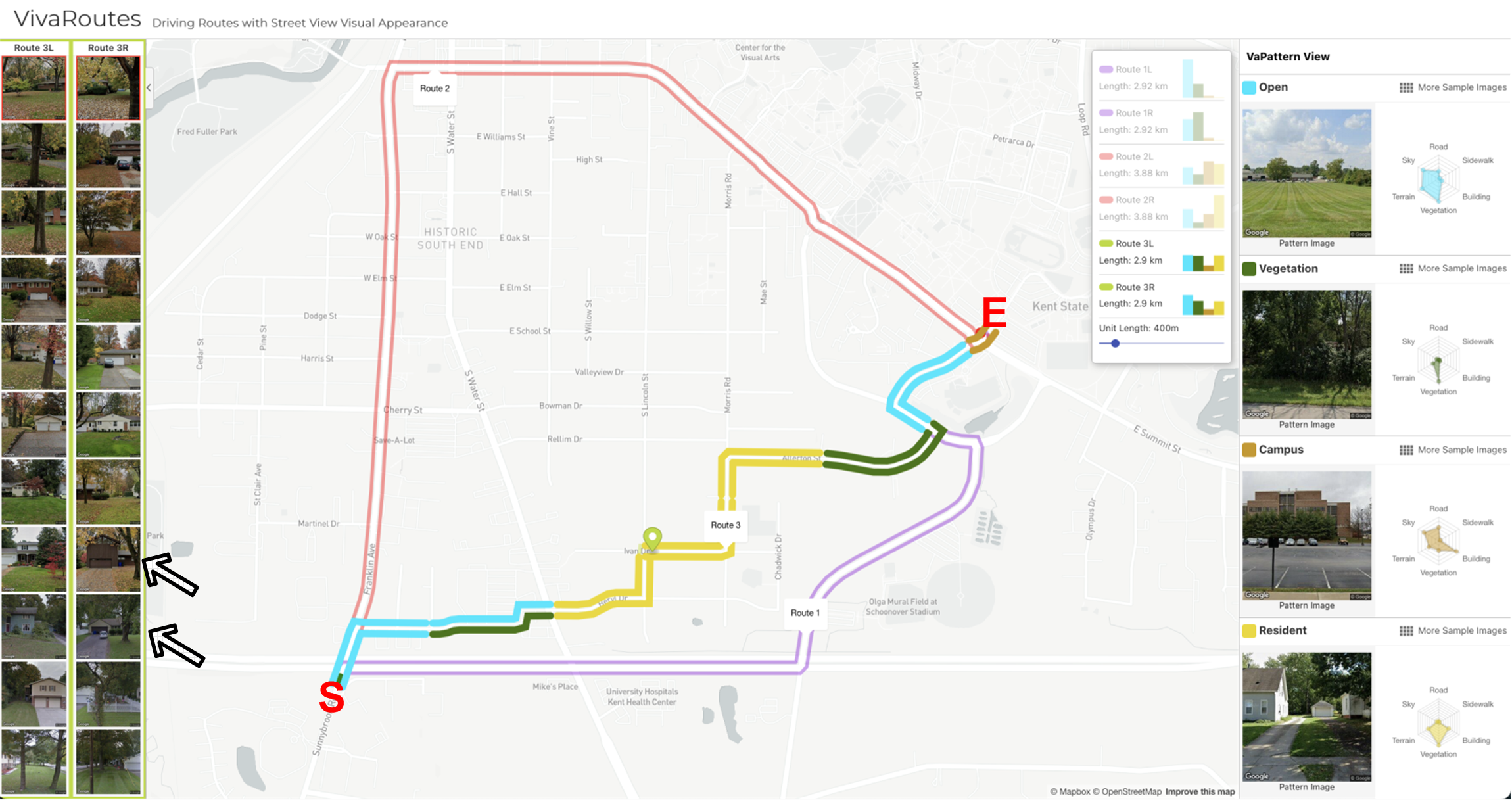}\vspace{-5pt}
\caption{Studying driving routes in a Midwest university town. The pattern image and a group of sample images (extracted from Google Street-View API) are shown together with the map view.} \vspace{-5pt}\label{fig:Kent1}
\end{figure*}

\section{Case Studies}
\label{sec:case_studies}
In this section, we use ViVaRoutes in two different geographical areas to show the usage scenarios.

\subsection{Understanding Visual Appearance Patterns in New York City}
VaPatterns provide abstracted city view patterns for quick visual exploration. As shown in Fig. \ref{fig:main_interface}{C} and in Fig. \ref{fig:VaCluster}, four VaPatterns are discovered from street-view images in Manhattan, New York. These VaPatterns reflect very unique visual appearance patterns of Manhattan. Users can edit and give the name of each pattern for easy description. Here in Fig. \ref{fig:VaCluster}, four patterns are shown side-by-side with the enlarged pattern images, categorical distribution charts, and groups of example representative images. In particular, the four patterns are:

\begin{itemize}
    \item Ordinary City Pattern (VaPattern 1): This pattern represents an ordinary street appearance with mixed buildings, sidewalks, bushes, and meadows. The typical semantic distribution of this pattern can be seen from the radar chart in Fig. \ref{fig:VaCluster}A: high-level of building with medium vegetation, and some road and sky. The pattern image in Fig. \ref{fig:VaCluster}{A} indicates this pattern’s visual appearance, along with a set of sample images below it.

    \item Infrastructure Pattern (VaPattern 2): This pattern includes the street-view images occupied mostly by buildings, bridges, and other transportation infrastructures. It can be seen from the radar chart in Fig. \ref{fig:VaCluster}{B} that on average, the dominant semantic category is building identified by the AI tool, while other categories are relatively negligible. From the sample images, it can be seen that these street-view images are largely dominated by infrastructure and lack the presence of the sky.

    \item Greenery Pattern (VaPattern 3): This pattern reflects the green visual appearance of the city with a large portion of trees, plants, and grasslands. It can be seen from Fig. \ref{fig:VaCluster}{C} that this pattern has a high percentage of vegetation in the chart, while the greenery images present the pattern to users.

    \item Open View Pattern (VaPattern 4): This pattern is about the city views with open sky or space, such as faraway views over water surfaces, empty fields, or squares. The radar chart in Fig. \ref{fig:VaCluster}{D} shows that this pattern is featured with open sky. The sample images show rivers with small buildings in a long distance.
\end{itemize}

These patterns summarize the city view features in Manhattan. Users can quickly recognize and perceive the basic visual appearances in this specific region from them. Therefore, this abstraction helps them explore multiple driving routes in map view with ease of understanding. 

\subsection{Studying Driving Routes with VaPatterns in New York City}
In this section, we present an example study where a user explores a city with the discovered patterns.

A user (named Alice for easy description) wants to travel from a starting location in Lower Manhattan to a destination in Upper Manhattan. She employs VivaRoutes to study the driving routes with an interest of street-views.  After she inputs the source and destination locations, VivaRoutes recommends three routes she can take on the map (Fig. \ref{fig:main_interface}{A}). Alice can click to select either route so that the route is visualized by colors encoding the street-view patterns. As illustrated in Fig. \ref{fig:main_interface}{A}, two alternative driving routes (Route 1 and 3) traverse along the western and eastern sides of Manhattan Island, respectively. They both use high-speed expressways to avoid going through small streets in the mid-town area. It is of great interest for Alice to find: “What are the different views when driving along the Hudson River (Route 1 on the western side) or the East River (Route 3 on the eastern side), respectively?”  From the visualization, it can be seen clearly that Route 1 has more Greenery Patterns shown in yellow color than Route 3. On the other hand, Route 3 has more Infrastructure Patterns, represented by purple color. 

Next, Alice checks the Route Comparison Panel (Fig. \ref{fig:main_interface}{E}) to compare the two routes in detail. She compares the ``Left'' and ``Right'' sides (based on her driving direction) of these routes. She sees that the left-side view of Route 1 has lots of yellow patterns from origin to destination. In contrast, on the right side, Alice will meet many infrastructure views (purple) at the beginning, and after a while, beautiful scenery views (yellow) will appear. On the other hand, on Route 3, infrastructure views (purple) can be seen on both the left and right sides. By dragging the marker on the routes, the sampled street-view images are shown in Fig. \ref{fig:main_interface}{A}. Alice quickly checks the views of interesting locations, such as the park view on Route 1 and the structures on Route 3.

This case shows that VivaRoutes can help Alice quickly form insights about her potential routes so that to make decisions accordingly. For example, she may choose Route 1 for the scene of the Hudson River, but she may also want to choose Route 3 for New York buildings and bridges.

\subsection{Exploring Driving Routes in a Midwest suburban town}

We further show the use of VivaRoutes in a Midwest region of Ohio. This suburban region has very different visual appearance features compared to New York City. Therefore, four different VaPatterns are discovered from its street-view images. As shown in the VaPattern View of Fig. \ref{fig:Kent1}, these patterns are named as:
\begin{itemize}
    \item Open Pattern: This pattern shows open grassland, sky, and road views.
    \item Vegetation Pattern: This pattern reflects views of trees and bushes.
    \item Campus Pattern: This pattern mostly includes views from a MidWest university campus in this suburban town.
    \item Resident Pattern: This pattern shows mostly the views of resident buildings and neighborhoods.
\end{itemize}
From them, we can recognize this area as a typical university town with mixed campus buildings, resident houses, trees, and grasses. Next, we show how Alice can use the visualization system for a study of multiple routes.

As shown in Fig. \ref{fig:Kent1}, Alice sets the starting location (S) as an off-campus resident building and the end location (E) as a university building on campus. VivaRoutes generates three recommended driving paths from S to E. She examines the map inlet to get a quick overview of these routes on each side. From the bar charts, Alice finds that Route 1 has mostly an open view and vegetation view. It does not include a lot of campus views and resident views. In comparison, Route 2 has a VaPattern distribution where campus views and resident views are prevalent. Route 3 has a more even distribution of the four patterns. Alice is interested in Route 3 as it offers more typical street views of this university town. She selects Route 3, and then the VaPatterns of 3L and 3R are highlighted on the map (Fig. \ref{fig:Kent1}).  Alice drags the marker along this route to examine the resident pattern views highlighted in yellow. On the image view, Alice finds many resident houses along this path. She may drive this route to get more inside information about this town. It can be seen that these images are not sampled in the same season (as indicated by the two arrows), but they are correctly grouped in the resident VaPattern. Similarly, Alice may choose Route 1 where she can find more green views and open views if she does not want to pass through the resident area.

\section{User Study}
\label{sec:userstudy}
We conducted user studies to evaluate our work in two major directions. First, we assessed whether visualizing visual appearance patterns for driving routes is acceptable to people and whether this goal is achieved by our computational and visualization techniques. Second, we evaluated the VivaRoutes system for its usability.  

\begin{table}[t]				
\caption{\it User Evaluation of Visualizing Visual Appearance Patterns}			\centering				
{\makebox{\resizebox{\columnwidth}{!}{				
\begin{tabular}{l|c|c} \hline%\noalign{\smallskip}		
\hspace{50pt} \textbf{Questions} & \textbf{Mean} & \textbf{Std. } \\	\hspace{30pt} 0(poor) - 10(excellent) &\textbf{Value} & \textbf{Dev.} \\
\hline%\noalign{\smallskip}				

Q1: Do you often use map-based navigation services &9.2 & 1.2\\		
 ~~(e.g. Google map, Apple map, etc.) when planning a new route? &&\\
 \hline
 
Q2: Is the new visualization of the streetview information &7.4& 1.5 \\
~~ potentially useful for you or others? && \\	
 \hline

Q3: How are you familiar with the geo-environment in this example? &7.6& 1.9 \\
\hline

Q4: Do you think the four patterns can represent the visual &8.0& 1.3 \\
~~ appearance of this area? &&\\
\hline

Q5: Do you agree that the pattern distribution shown on the three  &7.9&1.4\\
~~routes is reasonable, based on your own experience? &&\\
\hline

Q6: Can you quickly get the information from the visualizations? &7.9 &1.6 \\
\hline

Q7: Does the map view convey useful information? &8.1 &1.4\\
\hline

Q8: Does the RouteViewLine convey useful information? &7.4 &1.7\\
\hline

Q9: Does the coordinated raw street-view images convey &8.4 &1.4 \\
~~useful information? &&\\
\hline

\hline%\noalign{\smallskip}				
\end{tabular}				
}} } \vspace{-10pt}
\label{table:study}				
\end{table}

\begin{figure}
    \centering
 \includegraphics[width=0.8\columnwidth]{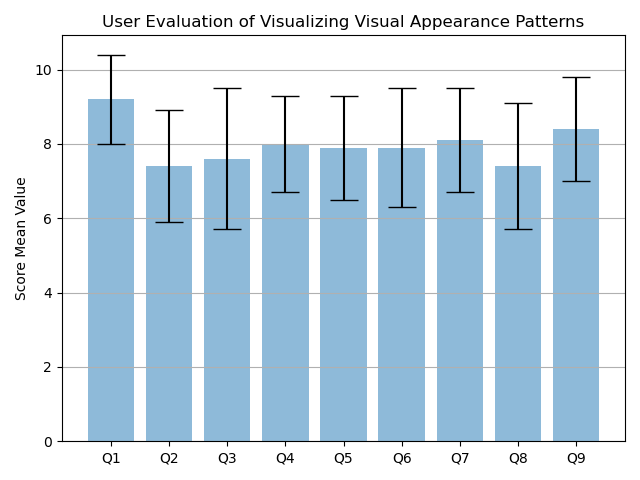}
\vspace{-5pt}
\caption{Bar chart view of User Evaluation Results in Table \ref{table:study}. }  \label{fig:barchart1}
\end{figure}

\subsection{User Study of Visualizing Visual Appearance Patterns}

\noindent \textbf{Participants}: We invited 45 participants (ages between 19 to 45, 21 males and 24 females) who are students living in a university town as shown in Fig. \ref{fig:Kent1}.  

\noindent \textbf{Procedure}: 
We developed an online VivaRoutes Demo System for the participants based on the data shown in Fig. \ref{fig:Kent1}. The geographical information and streetview images in the university town were incorporated, and the corresponding visual patterns were discovered and shown to the participants. 

In the user study, we scheduled individual interviews lasting 1 to 1.5 hours with each participant. We first discussed our motivation and introduced the work. Then, the participants were guided to explore the demo system. Finally, they were asked to fill in a form for a group of questions. These questions are shown in Table \ref{table:study} with the mean values and standard deviations from the participants' scores from 0 to 10. Fig. \ref{fig:barchart1} shows the bar chart view of the results. 

The following insights are derived from the results:

\begin{itemize}
    \item \textbf{Visualizing visual appearance patterns is useful}: As indicated in Table \ref{table:study}, the participants were familiar with map based route services (Q1 with mean 9.2). Then in Q2 (mean 7.4), they mostly agreed that adding street-view information is potentially useful. We further asked them to list the application areas for the usage. The results indicate: (1) Planning trips in a new place as a tourist (87\%); (2) Route planning in daily life (36\%); (3) City planning (38\%); (4) Exploring maps for fun or knowledge (67\%); (5) Community study (27\%). The results show that the participants perceived the new visual information as useful in various directions, primarily in route planning and exploration. It is worth mentioning that no one replied that the visual appearance patterns are not useful at all.

    \item \textbf{The visualization results is reasonable and meaningful}: Next, we asked the participants to evaluate the visual patterns. The answer to Q3 (mean 7.6) shows that they were mostly familiar with the geographical area and landscape environments, since this is the university town where most of them are living. Then for Q4 (mean 8.0), they confirmed that the four patterns we discovered by the algorithms (Sec. \ref{sec:patterndiscovery}) are meaningful based on their own knowledge of this town. Moreover, they examined the visualized patterns on the three candidate routes. In Q5 (mean 7.9), they agreed that the patterns shown on the routes are reasonable, which met their experiences on these routes.

    \item \textbf{The visualization is informative}: In Q6, they agreed that the visualizations convey the information in a good way (mean 7.9). Furthermore, they gave assessments for three major views. For the map view in Q7, the mean score is 8.1. For the RouteViewLine, the mean score of Q8 is 7.4. In addition, they liked the coordinated images in Q9 with a mean of 8.4. The scores indicate that the participants generally considered the visualizations informative.

\end{itemize}

\subsection{QUIS of VivaRoutes System}
We further conducted a study to evaluate the interactions in the VivaRoutes system. In this study, we invited 30 participants who are graduate students. Some of them have experience in visualization systems. They conducted this study one by one. 

\begin{figure}
    \centering
 \includegraphics[width=0.8\columnwidth]{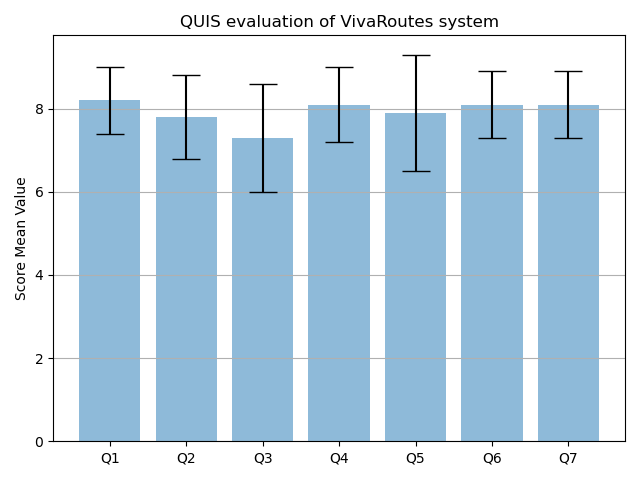}
\vspace{-5pt}
\caption{Bar chart view of QUIS Evaluation Results in Table \ref{table:QUIS}. }  \label{fig:barchart2}
\end{figure}

An instructor introduced the demo system, followed by explaining the visualization and interactions to each participant. They were guided to freely explore the system for about 15 minutes. Then, they defined routes and visualized the visual patterns. Afterward, they filled out a QUIS (Questionnaire for User Interaction Satisfaction) and provided written comments. Table \ref{table:QUIS} shows the questions and ratings. Fig. \ref{fig:barchart2} shows the bar chart view of the results. The average scores of all the questions are above 7.0, which is very good.

\subsection{User Feedback}
The participants commented the system is easy to operate and the system response time is very fast. Several participants indicated that the RouteViewLine  view may need more time for understanding. This may be attributed to the longer learning curve of this new design, which is different from the map view and image view that most people are familiar with. Some participants also pointed out that the visualizations in the current format may not be easily extended to mobile platforms. Since most people now use mobile phones for route planning, the system could be further enhanced for mobile applications. Moreover, it was commented that when traffic and other information need to be included, cluttering may happen. It could be a challenge since now the route color is used for visual appearance information. 

They also provided many suggestions, including reducing the amount of information presented to users, adding more labels and guidance in the system to shorten the learning curve, enlarging the photos for detailed study, hiding and popping up some parts for easy depiction, and integrating with satellite images, among others.

\begin{table}[t]				
\caption{\it QUIS evaluation of VivaRoutes system}			\centering				
{\makebox{\resizebox{\columnwidth}{!}{				
\begin{tabular}{l|c|c} \hline%\noalign{\smallskip}		
\hspace{50pt} \textbf{Questions} & \textbf{Mean} & \textbf{Std. } \\	\hspace{30pt}  &\textbf{Value} & \textbf{Dev.} \\
\hline%\noalign{\smallskip}				

Reading labels and icons on the screen. &8.2 & 0.8\\		
 ~~0(very hard)-9(very easy)  &&\\
 \hline
 
Selecting and highlighting items/areas.  &7.8& 1.0 \\
~~ 0(not at all)-9(very much) && \\	
 \hline

Organizing information in the interface with positions and layouts.  &7.3& 1.3 \\
~~ 0(confusing)-9(very clear) && \\	
\hline

Sequential operations on the interface. &8.1& 0.9 \\
~~ 0 (confusing)-9 (very clear) &&\\
\hline

Interaction on visual interface &7.9&1.4\\
~~0 (very hard) - 9 (very easy) &&\\
\hline

Learning to operate the system.  &8.1 &0.8 \\
~~ 0 (difficulty) - 9 (easy) &&\\
\hline

System response with good speed.  &8.1 &0.8 \\
~~0 (very slow) - 9 (fast enough) &&\\
\hline

\hline%\noalign{\smallskip}				
\end{tabular}				
}} } \vspace{-5pt}
\label{table:QUIS}				
\end{table}	

\section{Conclusion and Discussion}
\label{sec:conclusion}
In this study, we develop algorithms to discover visual appearance features along driving routes from the street-view imagery dataset. Visual semantic categories are extracted from the image with AI tools, which are also utilized to quantitatively cluster images to find visual appearance patterns. A VivaRoutes system presents the categories and patterns within an interactive geographical visualization interface, which allows users to explore the discovered features for route planning and decision-making. 

The approach identifies and presents street-view features based on the recommended routes from a route planner. These routes may be computed by time, cost, traffic, and other transportation factors. However, visual appearance features are not used in route computation. It is of great interest if routes can be recommended by utilizing these features, such as finding a route with mostly greenery views. In future work, we will incorporate visual semantic categories and/or VaPatterns in graph-based routing algorithms to further extend this work.

Furthermore, this work studies AI-based inductive approaches for discovering visual appearance patterns. We acknowledge that it is challenging to theoretically explain and analyze such patterns from diverse visual appearances. We will study more theoretical aspects of representing these street-view images in the future. In addition, city-level patterns may be an interesting topic for GIS applications. Another future direction is to extend this route-based work to city-level visual appearance extraction and visualization.

Finally, we will also improve VivaRoutes with enhanced visualization and interactions for public deployment.

\section*{Acknowledgement}
The street view images are extracted from Google Street View Static API. The map view uses Google Map API. The routes are acquired through Google Directions API. D3.js is used in generating visualizations.

% \newpage
\bibliographystyle{IEEEtran}      % IEEE

\bibliography{Driving}
\end{document}